\documentclass{iopart}
\usepackage{graphicx}
\usepackage{amssymb}
\usepackage{epstopdf,subfigure}

\begin{document}

\title{Zero-point energy in spheroidal geometries}
\author{A R Kitson and A I Signal}
\address{Institute of Fundamental Sciences, Massey University, Private Bag 11 222, Palmerston North, New Zealand}

\begin{abstract}
We study the zero-point energy of a massless scalar field subject to spheroidal boundary conditions. Using the zeta-function method, the zero-point energy is evaluated for small ellipticity. Axially symmetric vector fields are also considered. The results are interpreted within the context of QCD flux tubes and the MIT bag model.
\end{abstract}

\section{Introduction}
It is well known that the zero-point energy of a quantum field is highly dependent on the geometry of any boundary conditions. Many geometries have been studied including spherical, cylindrical and parallel plates~\cite{miltonbook}. These three geometries are the limiting cases of spheroidal geometry. By considering spheroidal geometry we may investigate mathematically how the zero-point energy changes as the geometry changes. In particular, the sign of the zero-point energy can be examined.

QCD is often studied phenomenologically in terms of bag models~\cite{mit}. The zero-point energy of a spherical bag has already been found~\cite{milton}. Lattice QCD simulations have shown that there exist flux tubes between quark anti-quark pairs and also three quark configurations~\cite{bissey}. We use spheroidal geometry to model these flux tubes. Using the infrared modified gluon bag model~\cite{oxman}, it is found that the zero-point energy produces an attractive inter-quark force.

\section{Spheroidal geometry}
The Cartesian equation for a spheroid is
\begin{equation}
\frac{x^{2}+y^{2}}{a^{2}}+\frac{z^{2}}{b^{2}}=1.
\end{equation}
If \(a<b\) (\(a>b\)), then the spheroid is prolate (oblate). If \(a=b\), then the spheroid is a sphere of radius $a$. Prolate spheroidal curvilinear coordinates, \((u, v, w)\), are related to Cartesian coordinates by
\begin{eqnarray}
x&=&f\sqrt{(u^{2}-1)(1-v^{2})}\cos(w)\\
y&=&f\sqrt{(u^{2}-1)(1-v^{2})}\sin(w)\\
z&=&f u v
\end{eqnarray}
where \(f=\sqrt{b^{2}-a^{2}}\) and \(1\leq u<\infty,-1\leq v\le1,0\leq w<2\pi\). In these coordinates, surfaces of constant $u$ are confocal prolate spheroids. The prolate ellipticity, $e$, is a measure of deviation from a sphere and is defined by \(e=f/b\in[0,1)\).

\section{Scalar field}

\subsection{Spheroidal functions}
We consider a real, massless scalar field, $\phi$, confined within a prolate spheroid with Dirichlet boundary conditions. The field equation (with \(\hbar=c=1\)) and boundary condition are
\begin{eqnarray}
\left(\nabla^{2}+\omega^{2}\right)\phi=0,\label{helmholtz}\\
\phi|_{u=1/e}=0,
\end{eqnarray}
where $\omega$ is the frequency. It can be shown that~\eref{helmholtz} is separable in prolate spheroidal coordinates~\cite{morse}. Let \(\phi=U(u)V(v)W(w)\), then~\eref{helmholtz} separates into three ordinary differential equations
\begin{eqnarray}
\left(\frac{\rmd}{\rmd u}\left((u^{2}-1)\frac{\rmd}{\rmd u}\right)-\lambda+c^{2}u^{2}+\frac{\mu^{2}}{u^{2}-1}\right)U(u)=0,\label{U}\\
\left(\frac{\rmd}{\rmd v}\left((1-v^{2})\frac{\rmd}{\rmd v}\right)+\lambda-c^{2}v^{2}-\frac{\mu^{2}}{1-v^{2}}\right)V(v)=0\label{V},\\
\left(\frac{\rmd^{2}}{\rmd w^{2}}+\mu^{2}\right)W(w)=0,\label{W}
\end{eqnarray}
where \(c=\omega f\) and $\mu$ and $\lambda$ are separation constants. The solution to~\eref{W} is straightforward and requires \(\mu=m\in\mathbb{Z}\). The solutions to~\eref{U} and~\eref{V} that are regular and satisfy the boundary condition are the radial spheroidal function of the first kind, \(R_{m,n}^{(1)}(c,u)\), and the angular spheroidal function of the first kind, \(S_{m,n}^{(1)}(c,v)\), respectively. Here $n$ is an integer such that \(n\ge|m|\). In the spherical limit,~\eref{U} reduces to the spherical Bessel differential equation and~\eref{V} reduces to the associated Legendre differential equation. In fact, the radial (angular) spheroidal functions can be expanded in a basis of spherical Bessel (associated Legendre) functions~\cite{falloon}
\begin{eqnarray}
R_{m,n}^{(1)}(c,u)=\frac{(u^{2}-1)^{m/2}}{u^{m}} \sum_{k=-\infty}^{\infty} a_{2k}^{m,n}(c)j_{n+2k}(c u), \label{radialspheroidal}\\
S_{m,n}^{(1)}(c,v)=\sum_{k=-\infty}^{\infty} d_{2k}^{m,n}(c)P_{n+2k}^{m}(v),
\end{eqnarray}
where the coefficients $a_{2k}^{m,n}(c)$ and $d_{2k}^{m,n}(c)$ satisfy certain three-term recurrence relations. Angular spheroidal functions are normalized in the same fashion as associate Legendre functions, whereas the radial spheroidal functions are normalized to have the same asymptotic form as spherical Bessel functions
\begin{equation}
R_{m,n}^{(1)}(c,u)\stackrel{c u\rightarrow \infty}{\sim}j_{n}(c u).
\end{equation}
The Dirichlet boundary condition can now be written in terms of radial spheroidal functions
\begin{equation}
R_{m,n}^{(1)}(z_{m,n,k} e,1/e)=0,\label{eigenfreq}
\end{equation}
where \(z_{m,n,k}=\omega_{m,n,k} b\) and $\omega_{m,n,k}$ is the $k^{th}$ eigen-frequency. Note that  $b$ and $e$ are used for the geometrical variables however, since only small ellipticity will be considered $a$ and $e$ could have equally been used.

\subsection{Zero-point energy}
In terms of the eigen-frequencies of a quantum field subject to some boundary condition, the zero-point energy is
\begin{equation}
E=\frac{1}{2}\sum_{m,n,k} \omega_{m,n,k}.
\end{equation}
As expressed above, the zero-point energy is divergent and must be regularized. The prolate spheroidal zeta function is defined
\begin{equation}
\zeta_{m,n}(s)=(b \mu)^{s}\sum_{k=1}^{\infty} z_{m,n,k}^{-s},\label{pszeta}
\end{equation}
where the $z_{m,n,k}$ come from~\eref{eigenfreq}, $\mu$ is an arbitrary scale with mass dimensions and the dependance on $e$ has been suppressed. As written,~\eref{pszeta} is only valid for \({\rm Re}\, s>1\). The regularized zero-point energy is defined in terms of a principal part prescription~\cite{blau},
\begin{equation}
E\equiv\frac{1}{2}\mu{\bf PP}_{s\rightarrow -1}\left[\sum_{m,n} \zeta_{m,n} (s)\right]. \label{principlepart}
\end{equation}
Clearly, knowledge of the prolate spheroidal zeta function as \(s\rightarrow -1\) is required. Fortunately, $\zeta(s)$ may be analytically continued to the region \({-1<\rm Re}\, s<0\). Using the argument principal,~\eref{pszeta} may be re-written
\begin{equation}
\zeta_{m,n}(s)=(b \mu)^{s} \frac{s}{2 \pi i}\int_{C} \rmd z\,z^{-s-1} \ln(R_{m,n}^{(1)}(z e,1/e)),
\end{equation}
where the contour encloses all the zeros of $R_{m,n}^{(1)}(z e,1/e)$. Because the radial spheroidal functions have the same asymptotic form as the spherical Bessel functions, we can follow an analogous argument to Romeo~\cite{romeoscalar}, arriving at an integral expression for the spheroidal zeta function valid for \({-1<\rm Re}\, s<0\),
\begin{equation}
\fl
\zeta_{m,n}(s)=(b \mu)^{s}\frac{s}{\pi}\sin\left(\frac{\pi s}{2}\right)\int_{0}^{\infty}\rmd x\,x^{-s-1}\ln\left(2 x \exp(-x) T_{m,n}^{(1)}(x e,1/e)\right),
\end{equation}
where $T_{m,n}^{(1)}(x e,1/e)$ are the modified radial prolate spheroidal functions, modified in the same fashion as the modified spherical Bessel functions. If we make a Taylor expansion for small ellipticity, then
\begin{eqnarray}
\fl
\lefteqn{\sum_{m,n}\zeta_{m,n}(s) = (b \mu)^{s}\frac{s}{\pi}\sin\left(\frac{\pi s}{2}\right)\sum_{n=0}^{\infty}2\nu^{s+1}\int_{0}^{\infty}\rmd x\,x^{-s-1}\ln\left(2 \exp(-\nu x) s_{n}(\nu x)\right)}\nonumber\\
\fl
+\left[(b \mu)^{s}\frac{s}{\pi}\sin\left(\frac{\pi s}{2}\right)\sum_{n=0}^{\infty}\nu^{s}\int_{0}^{\infty}\rmd x\,x^{-s-1}\frac{2 \nu}{3}\left(1-\frac{x s_{n}^{\prime}(\nu x)}{s_{n}(\nu x)}\right)\right]e^{2}+\Or\left(e^{4}\right),\label{sumzeta} 
\end{eqnarray}
where \(\nu=n+1/2\) and $s_{n}(\nu x)$ are the Riccati-Bessel functions of imaginary argument. The first term in~\eref{sumzeta} is simply the spherical zeta function and has already been evaluated~\cite{romeoscalar}; the second term is new and represents the first order correction to the spheroidal zeta function for small ellipticity.  To evaluate the second term, the standard procedure is followed. Using the uniform asymptotic expansion of the modified spherical Bessel functions~\cite{abramowitz}, the asymptotic form of the integrand is added and subtracted. The integrand is expanded to $\Or\left(\nu^{-5}\right)$, which is the same order as in~\cite{romeoscalar}. The integral of the added terms can be expressed in terms of beta functions and the integral of the subtracted terms can be evaluated numerically. The sums over $n$ can be expressed in terms of Riemann zeta functions. The result, when Laurent expanded around \(s=-1\), is
\begin{eqnarray}
\fl
\sum_{m,n} \zeta(s)=\frac{1}{b \mu}\left(\frac{2}{315 \pi}\left(\frac{1}{s+1}+\ln(b \mu)\right)+0.0089\right)+\Or\left(s+1\right)\nonumber\\
\fl
+\left[\frac{1}{b \mu}\left(\frac{2}{945 \pi}\left(\frac{1}{s+1}+\ln(b \mu)\right)+0.0023\right)+\Or\left(s+1\right)\right]e^{2}+\Or\left(e^{4}\right).
\end{eqnarray}
Thus, using~\eref{principlepart},
\begin{eqnarray}
\fl
E^{\rm I}(\mu)=\frac{1}{2 b}\left(\frac{2}{315 \pi}\ln(b \mu)+0.0089\right)\nonumber\\
+\left[\frac{1}{2 b}\left(\frac{2}{945 \pi}\ln(b \mu)+0.0023\right)\right]e^{2}+\Or\left(e^{4}\right),
\end{eqnarray}
where the superscript, ${\rm I}$, signifies that this is the zero-point energy for internal eigen-frequencies only. 

If the field is confined to the outside of a prolate spheroid by Dirichlet boundary conditions, then the exterior eigen-frequencies are found. The exterior zero-point energy can be calculated using an analogous spheroidal zeta function. If internal and external eigen-frequencies are considered together, then the total zero-point energy is
\begin{equation}
E_{\rm prolate}=\frac{1}{2 b}\left(0.0056+0.0019 e^{2}\right)+\Or\left(e^{4}\right).\label{pstotal}
\end{equation}
As in the spherical case, there is no pole at \(s=-1\) in the total prolate spheroidal zeta function, so the principal part prescription is not needed. The total zero-point energy is independent of the mass scale.

The zero-point energy of a real, massless scalar field subject to the same boundary conditions but on the surface of an oblate spheroid has also been calculated~\cite{kitson}. We quote here the result for the total energy
\begin{equation}
E_{\rm oblate}=\frac{1}{2 b}\left(0.0056-0.0019 {\tilde{e}}^{2}\right)+\Or\left({\tilde{e}}^{4}\right),\label{ostotal}
\end{equation}
where \(\tilde{e}^{2}=a^{2}/b^{2}-1\) and is related to the oblate ellipticity. In both~(\ref{pstotal}) and~(\ref{ostotal}) the spherical part agrees with~\cite{bender}.

\subsection{Green's functions}
The results of the previous section can be confirmed using the Green's function approach. If all space is considered, then the zero-point energy is given by~\cite{schwinger}
\begin{equation}
E\equiv\lim_{x^{\prime}\rightarrow x} \frac{1}{2 \rmi}\int \frac{\rmd\omega}{2\pi}\,e^{-\rmi \omega \left(t-t^{\prime}\right)}\int \rmd\bi{r}\,2 \omega^{2} g\left(\bi{r},\bi{r^{\prime}}\right),\label{energygreens}
\end{equation}
where $g\left(\bi{r},\bi{r^{\prime}}\right)$ is the reduced Green's function, which satisfies
\begin{equation}
\left(\bi{\nabla}^{2}+\omega^{2}\right) g\left(\bi{r},\bi{r^{\prime}}\right)=-\delta(\bi{r}-\bi{r^{\prime}}).
\end{equation}
The $g\left(\bi{r},\bi{r^{\prime}}\right)$ that satisfies the Dirichlet boundary condition on the surface of a prolate spheroid is
\begin{eqnarray}
\fl
g\left(\bi{r},\bi{r^{\prime}}\right)=-\rmi \omega\sum_{m,n}X_{m,n}(c,v,w)X_{m,n}^{*}(c,v^{\prime},w^{\prime})\nonumber\\
\times\left\{\begin{array}{ll}
A R_{m,n}^{(1)}(c, u)R_{m,n}^{(1)}(c, u^{\prime})\qquad&u,u^{\prime}<1/e\\
B R_{m,n}^{(3)}(c, u)R_{m,n}^{(3)}(c, u^{\prime})&u,u^{\prime}>1/e\end{array}\right.,                                                                                                                                                                                                                                                                                                                                                                             
\end{eqnarray}
where the coefficients are
\begin{equation}
A=B^{-1}=\frac{R_{m,n}^{(3)}(c, 1/e)}{R_{m,n}^{(1)}(c, 1/e)},
\end{equation}
and $R_{m,n}^{(3)}(c, u)$ are the radial prolate spheroidal functions of the third kind, which are given by equation~(\ref{radialspheroidal}) with spherical Hankel functions of the first kind, $h_{n}^{(1)}(c u)$, as the basis functions (instead of $j_{n}(c u)$). The prolate spheroidal harmonics are
\begin{equation}
X_{m,n}(c,v,w)=\sqrt{\frac{2n+1}{4\pi}\frac{(n-m)!}{(n+m)!}}S_{m,n}^{(1)}(c,v)e^{i m w}.
\end{equation}
If we expand~(\ref{energygreens}) for small ellipticity and expand the resulting integrands asymptotically to $\Or\left(\nu^{-2}\right)$ the result agrees with~\eref{pstotal} to four decimal places.

\subsection{Results}
\begin{figure}
\centering
\subfigure[Prolate spheroid]{
\label{fig1a}
\includegraphics[width=5.5cm]{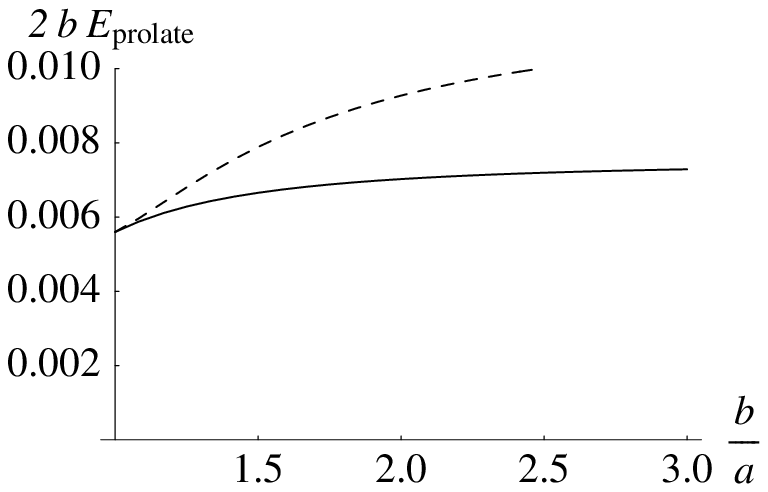}}
\hspace{1cm}
\subfigure[Oblate spheroid]{
\label{fig1b}
\includegraphics[width=5.5cm]{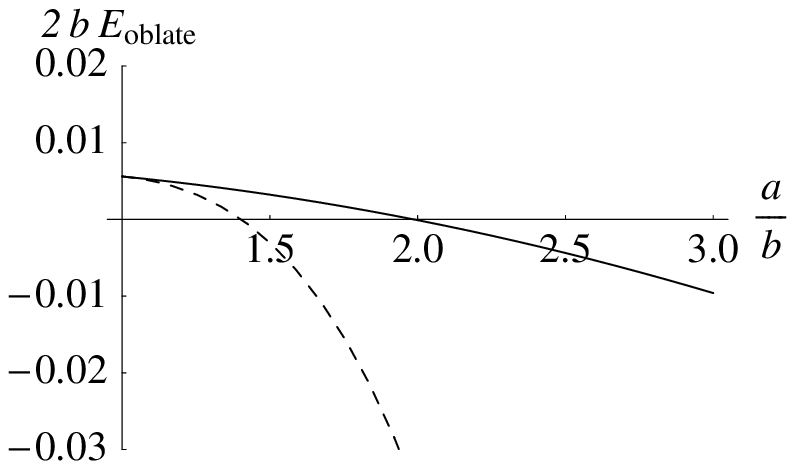}}
\caption{Total zero-point energy of a scalar field subject to Dirichlet conditions on a prolate spheroidal (a) and an oblate spheroid (b) of small ellipticity. The dashed lines represent preliminary $\Or\left(e^{4}\right)$ results.}
\label{plots}
\end{figure}

The results are shown in~\Fref{plots}. The preliminary $\Or\left(e^{4}\right)$ results (dashed lines) reinforce that~\eref{pstotal} and~\eref{ostotal} are only valid for small ellipticity.

It can be seen that $2 b E_{\rm prolate}$ increases as the sphere is deformed. This is expected as the limiting geometry is an infinite circular cylinder, for which it is known that the zero-point energy per unit length is positive~\cite{nesterenko}.

In the oblate case, the zero-point energy becomes negative. Again, this is expected since the zero-point energy per unit area of two parallel plates is negative~\cite{miltonbook}. The precise point at which the zero-point energy is zero is outside the region where these results are valid. Work is currently being done to find this point.

\section{Vector field}
The vector Helmholtz equation cannot be solved by separation of variables in prolate spheroidal coordinates~\cite{morse}. However, if the vector field is axially symmetric, then it can be separated into TE and TM modes. The TE and TM boundary conditions for an axially symmetric field confined within a perfectly conducting prolate spheroid cavity are, respectively
\begin{eqnarray}
\left.R_{1,n}^{(1)}(c, u)\right|_{u=1/e}=0, \\
\left.\frac{\partial}{\partial u}\left(\sqrt{u^{2}-1}\,R_{1,n}^{(1)}(c,u)\right)\right|_{u=1/e}=0.
\end{eqnarray}
It should be noted that, due to the assumed axial symmetry, the radial spheroidal functions have \(m=1\) and now \(n\in\mathbb{Z}^{+}\). Adding the contributions from both the TE and TM modes and using~\eref{principlepart}, the internal zero-point energy of an axially symmetric vector field is~\cite{kitson}
\begin{equation}
\fl
E_{\rm V}^{\rm I}(\mu)=\frac{1}{2 b}\left(\frac{16}{315 \pi}\ln(b \mu)+0.168\right)+\left[\frac{1}{2 b}\left(0.037\ln(b \mu)-0.057\right)\right]e^{2}+\Or\left(e^{4}\right).\label{internalvector}
\end{equation}
In this calculation the integrands were expanded asymptotically to $\Or\left(\nu^{-2}\right)$. The spherical part agrees with~\cite{romeovector}.

\subsection{QCD}
Assuming the infrared modified gluon bag model~\cite{oxman}, the internal zero-point energy of a colour field is approximately
\begin{equation}
E_{\rm QCD}^{\rm I}\approx-8E_{\rm V}^{\rm I},\label{qcdzpe}
\end{equation}
where the factor of $8$ comes from the eight gluon fields. The minus sign comes from modifying the free gluon propagator to one which vanishes in the infrared region.

We model an inter-quark flux tube by a prolate spheroidal cavity with bag-like boundary conditions. Using~\eref{qcdzpe}, the QCD zero-point energy is approximately
\begin{equation}
E_{\rm QCD}^{\rm I}\approx\frac{1}{2 b}\left(-1.34+0.45e^{2}\right)+\Or\left(e^{4}\right),\label{qcd}
\end{equation}
where the scale independent part of~\eref{internalvector} has been used and thus we must assume the colour fields are axially symmetric. If the flux tube is stretched, then~\eref{qcd} implies that the zero-point energy increases. Therefore, the zero-point energy contributes an attractive inter-quark force. The result also suggests that spherical bags are stable against deformations.

\ack
Massey University and the Royal Society of New Zealand are acknowledged for their financial support.

\end{document}